\documentclass[twocolumn,showpacs,preprintnumbers,amsmath,amssymb]{revtex4}

\usepackage{graphicx}

\begin{document} 

\title{Mobility and Social Network Effects on Extremist Opinions}


\author{Andr\'e C. R. Martins}
 \email{amartins@usp.br}
 \affiliation{%
GRIFE - EACH, Universidade de S\~ao Paulo\\
Av. Arlindo B\'etio, 1000, S\~ao Paulo, 03828-080, Brazil}


\date{\today}




\begin{abstract}
    Understanding the emergence of  extreme opinions  and in what kind of environment they might become less extreme is  a central theme in our modern globalized society. A  model combining continuous opinions and observed discrete actions (CODA) capable of addressing the important issue of measuring how extreme opinions might be has been recently proposed. In this paper I show extreme opinions to arise in a ubiquitous manner in the CODA model for a multitude of social network structures. Depending on network details reducing extremism seems to be possible. However, a large number agents with extreme opinions is always observed. A significant decrease in the number of extremists can be observed by allowing agents to change their positions in the network.
\end{abstract}

\maketitle 


\section{Introduction}

People with extreme opinions about some issue are not rarely observed around us. Depending on the content of those opinions, some people might feel justified into committing violent actions, including terrorism~\cite{atran03}. Therefore, a model that allows us to observe the emergence of extremist opinions can be an invaluable tool, both from a descriptive as well as from a practical and political point of view. Opinion Dynamics~\cite{bordognaalbano07,castellanoetal07} is an attempt at modeling the behavior of societies of interacting agents that change their opinions under the influence of other agents. The opinions of each agent are described either as discrete~\cite{galametal82,galammoscovici91,sznajd00,stauffer03a} or as continuous~\cite{deffuantetal00,hegselmannkrause02} variables. Each agent, after observing the opinion of one or more of its neighbors, changes its own mind by following a simplified set of rules. In previous models, extremism was observed only as lack of influence between groups ~\cite{deffuantetal00,hegselmannkrause02,galam05,amblarddeffuant04,deffuant06}. 

Discrete opinion models, where each agent can have only one of a finite number of different opinions may seem, at first, adequate at describing extreme behavior when two opposing world views are fighting. However, they don't allow for each agent to have very strong beliefs.  Frequently, there are two opposing points of view (different opinions represented usually by $s_i = \pm 1$). The opinion of each agent is updated by pre-defined rules, such as the voter model~\cite{cliffordsudbury73,holleyliggett75} - where a randomly drawn agent adopts the opinion of one of its neighbors - or the Sznajd model~\cite{sznajd00,stauffer03a} - where  the agreement between two neighbors is needed to influence the entire neighborhood. But discrete opinions have no strength associated with them and the only traditional way to study extremism is by introduing inflexible agents in the problem~\cite{galamjacobs07}.

Continuous opinion models, on the other hand, allow opinions to be extreme and, possibly, even divergent~\cite{deffuantetal00,hegselmannkrause02,bennaimetal03}. While the appearance of clusters that, after a while, stop influencing each other is observed, the identification of these clusters of opinions as extreme would be based only on this lack of influence. The actual value of those opinions is not necessarily extreme and, therefore, one should not identify those groups as extremists. The more natural way of defining extremism in those models is by indentifying those individuals whose numerical value of opinion corresponds to the more distant values (usually, but not necessarily, close to 0 or 1). However, since continuous opinion models are built with a dynamics where opinions only converge or do not change, the real extreme opinions must be introduced in the initial conditions of the models. Those characteristics make those continuous models good for describing the spread of extremism when extremism is already present. The effect of different networks or inflexible agents on the group behavior can be studied~\cite{weisbuchetal05,deffuant06}, but the inflexibility of agents must be artificially imposed and is not observed as a consequence of the model. A model that can allow extremism to be observed as a consequence of the dynamics is obviously more useful for understanding extremists.

By adopting a probabilistic description of the opinions, the Continuous Opinions and observed Discrete Actions (CODA) model, that I introduced in a previous paper~\cite{martins08a},  was able to describe the problem of opinion dynamics in terms that extreme opinions can actually be measured by how close each agent is to certainty. This was implemented by the introduction of Bayesian rules for the continuous opinion update. Therefore, an extremist can be defined as an agent who  supports one choice fervently, even when a large group believes a different idea to be a better choice. In this article, only opinions and their verbalization will be studied and not actions based on those opinions. However, since the more certain you are the more likely it is that you will act on your beliefs, the certainty should eventually translate into action. The effects of different networks on the amount of extremism that emerges from the model will be studied, as well as the consequences of introducing random mobility for the agents.

\section{The CODA model}

The CODA model distinguishes between opinion, measured as a continuous variable, and action, as a binary one. Such a distinction has been proposed earlier in discrete models~\cite{urbig03}, related to verbalization problems, in the context of a game bet. The idea is that sometimes opinions should be hard to change, while other opinions might carry less weight. Similar concepts have also been explored in the context of the voter model, by introducing an opinion inertia where the number of interactions needed before an opinion is changed increases with the time the opinion is held~\cite{starketal08a,starketal08b}. Interestingly, it was observed that this change, that slows agent dynamics, can speed the macro behavior of the system. 

In the CODA model, on the other hand, the opinions  of each agent $i$ are described as a continuous probability function $p_i$. Each agent assigns $p_i$ to the statement that one of two choices (or actions, $s_i = \pm 1$, for agent $i$) is the best. Agents never observe the value of $p_i$ of other agents, only the choices made by agents they interact with, $s_j=sign(p_j)$. This means that some observed opinions might be easy to change (when $p$ is close to 0.5), while others might require many interactions. After observing other agents choices, each agent update its own probabilities using a simple Bayesian calculation. The problem is more easily described in terms of the log-odds variable 
\begin{equation}~\label{eq:update}
v_i=\ln\left( \frac{p_i}{1-p_i} \right)
\end{equation}

Whenever an agent observes someone whose choice is $s_i =-1$, the Bayesian update rule shows that $v_i$ is changed to $v_i-a$, where $a$ depends on how likely the agent thinks it is that a neighbor will favor the right choice. Symmetrically, if $s_i =+1$ is observed, instead, $v_i$ is altered to $v_i+a$. The parameter $a$ can be obtained from the probability $\alpha=P(OA|A)$ each agent assigns to observing a neighbor $OA$ that chooses alternative $A$, as the best alternative when $A$ is indeed the best one and can be estimated by $a=\ln \frac{\alpha}{1-\alpha}$. That is, if one choose a conservative value for $\alpha$ such as 0.55, we have $a\approx 0.2$. However, its exact value makes little difference in interpreting the results. Since the movement in both directions is of the same size, the number of interactions needed to flip an opinion that drifted 100 steps away from changing will be 100 steps back, regardless of the size of the step. In all the simulations presented bellow, $\alpha$ was chosen to be the same for every agent.

 Equation~\ref{eq:update} is an invertible function; therefore, at any point, $v_i$ can be translated back to $p_i$. From here on, we will measure the opinions as the number of steps of size $a$ an agent is from the central opinion ($p_i=0.5$, or $v_i=0$). An agent changes binary opinion is trivially given by the option he believes more likely to be true, that is, for $v_i< 0$, $s_i =-1$ and $s_i =+1$, otherwise. This particular rule for changing $p_i$ can be used in conjunction with any pattern of interaction among agents (e.g. voter or the Sznajd models). CODA can also be implemented using any social network structure, from regular lattices to small worlds and random graphs~\cite{newmanetal02}. It is worth mentioning that, while in the social inertia model, consensus was achieved faster when the agents changed their opinions slower, things are different for the CODA model. As soon as small domains are formed, they reinforce each other in a way that they soon become very strong and very unlikely to change.
 
\section{Effects of Different Networks on Extremism}
 
It was observed that extremism emerges naturally in the CODA model, even when every agent starts with a moderate opinion~\cite{martins08a}. We can define extremism for one single agent, meaning his opinion is so strong that even after many influences in the contrary direction, that agent will still keep his choice. However, we are more interested in social phenomena and, therefore, the word extremism will be used, from now on, in a social sense. More exactly, a society will be called extremist when both choices survive in the long run and most of the agents that support either choice are extremists. If only one choice is observed to survive in the long run, the final status of the society will be called a consensus, even if the opinions that support that consensus are extreme (they will always be when a consensus emerges).
 
\begin{figure}[bbb]
 \includegraphics[width=0.5\textwidth]{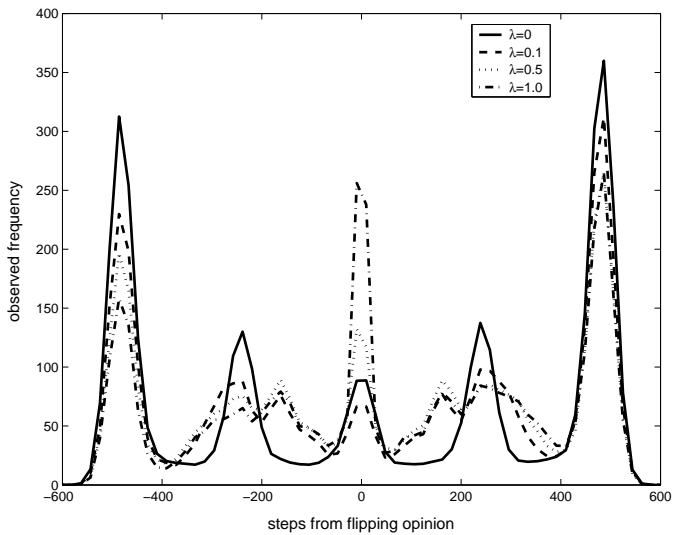}
 \label{fig:voter}
 \caption{Average distribution of the opinions of agents in a 64x64 square lattice after 2,000,000 of opinion updates (about 488 average updates per agent). The figure shows the cases for the standard voter update rule. The simulations show the results for a regular square lattice with 4 neighbors, as well as for small world lattices, where each link was randomly replaced with probability $\lambda$.}
\end{figure}

Figures~1 to 4 show simulation results for a community with 64x64 agents with several network structures.  The results were averaged over 20 different realizations of the dynamics and the curves correspond to the state of the system after $2\times 10^6$ updates (an average of about 488 updates per agent).  They show the observed frequency as a function of the number of steps each agent is distant from changing opinion. The positive side is chosen at each run as the side the majority supports and that is the origin of the asymmetry in the observed curves. In all cases shown, the update rule used was that of the voter model, where one voter and its influencing neighbor were randomly drawn at each iteration, with the first agent changing its continuous opinion towards that of the neighbor. In some of the cases, Sznajd model update rules have been employed. Having produced similar results, they are not shown here.

\begin{figure}[bbb]
 \includegraphics[width=0.5\textwidth]{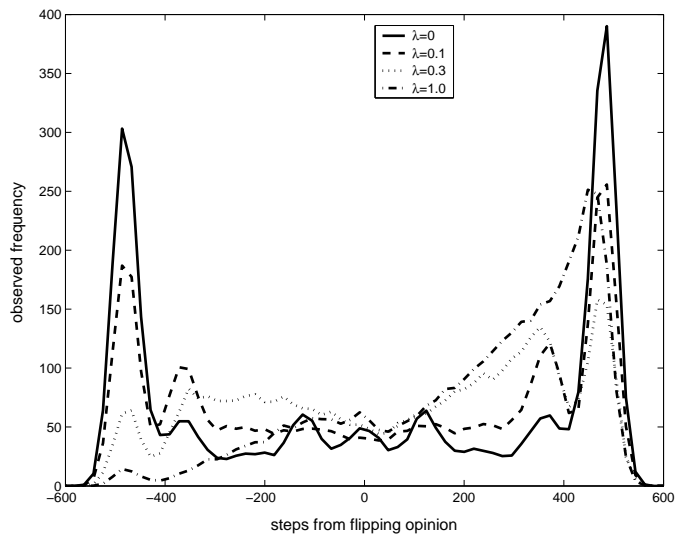}
 \label{fig:distributions}
 \caption{Average distribution of the opinions of agents in a 64x64 square lattice after 2,000,000 of opinion updates (about 488 average updates per agent). The simulations show the results for a regular square lattice with 8 neighbors as well as for small world lattices, where each link was randomly replaced with probability $\lambda$.}
\end{figure}

In all the Figures, the distance to the central opinion is measured in multiples of the step $a$, as discussed above. Therefore, it corresponds to the number of steps an agent is away from changing its opinion. Initial conditions were always chosen so that no extremists were present; each agent had a 50\% chance of supporting each opinion and all $v_i$ were drawn close enough to zero that the agent would change its opinion by meeting only one another agent who disagreed with him. The first three Figures correspond to alternative network structures. Figure~1 corresponds to a regular lattice with four first neighbors only, that is, a von Neumann neighborhood; Figure~2 includes diagonal vertices as neighbors (a Moore neighborhood). 
Figure~3 corresponds to the existence of two extra neighbors to the von Neumann neighborhood. Those extra agents do not change their opinions (influential neighbors), and each one favors a different, opposite choice (as an external constant field). In all of these first three cases, to investigate small-world effects, new simulations were also prepared where each link of the square lattice was altered with probability $\lambda$  and the curves corresponding to different values of $\lambda$ are presented for each case.

 \begin{figure}[bbb]
 \includegraphics[width=0.5\textwidth]{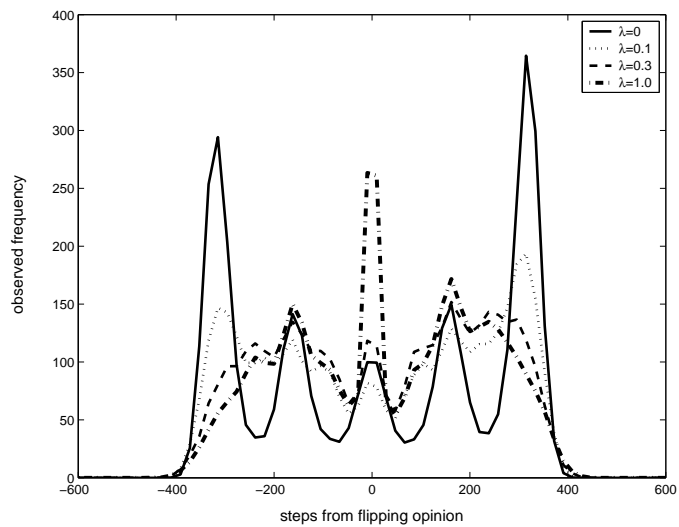}
 \label{fig:influential}
 \caption{Average distribution of the opinions of agents in a 64x64 square lattice after 2,000,000 of opinion updates (about 488 average updates per agent). The simulations show the results for a regular square lattice with 4 neighbors with two extra links added to each agent to two unchanging and opposite agents.  Results for small world lattices are also shown, where each link was randomly replaced with probability $\lambda$.}
\end{figure}

We can observe that, in all three cases, for regular lattices ($\lambda=0$), there are two large peaks at the most extreme opinions. 
Although extreme peaks are generally resistant to the introduction of randomness into the network connection pattern, we can see that small worlds have fewer extremist agents than regular lattices. This happens because once local neighborhoods of agreeing agents are formed their opinion is mutually reinforced.  Agents at boundaries will tend to agree with the majority of their neighbors in the long run without becoming extremists. This suggests that increasing the connectivity at the opinion boundaries, so that the majorities become less important, would weaken extremist tendencies.

 This idea can be tested comparing the four neighbors simulations with those were run where the agents were also influenced by diagonal neighbors. If the transition between opinions is a straight wall, in a traditional first-neighbor lattice, each agent is surrounded by 3 agents that think alike and only one, at the other side of the wall, who disagrees. This means that 3 in 4 times the agent change its opinions, it will be toward the opinion it already had. By introducing diagonal neighbors, the 3-to-1 proportion is changed to 5-to-3, allowing for less extremism in the borders. The results of those simulations can be seen in Figure~2. The effect for regular lattices is small and for lattices tending to a random graph (as $\lambda$ grows), one sees the appearance of a clear majority, with only a few agents having the opposite extreme opinion. There is an intermediary region, around $\lambda=0.3$, where some of the most extreme opinions are replaced by weaker, but still extreme opinions, but centrists are not more common than before. The problem is that, although agents in the borders tend to be less extreme, their opinions still change, in average, in the same direction and, in the long run, the diagonals have little impact inside the domains. And we can observe that $\lambda$ has to be too large for a clear majority to appear. It is important to notice that clear majorities and consensus still correspond to cases where most agents have extreme opinions, but, under those circumstances, most of the society would agree they are correct.

The influential neighbors case (without diagonal neighbors) shows that when everyone is influenced by both points of view, the extreme positions become a little less extreme (notice that the distribution of opinions is shrunk to an area a little closer to $v=0$).  The extremist peak, after the same number of interactions, is located close to 300 steps away from 0, instead of around 500 steps, as it was in the simulations with no influential neighbors. This is still extremism, though, as it corresponds to values of $p$ too close to certainty.

\begin{figure}[bbb]\label{fig:diagonallam1}
 \includegraphics[width=0.45\textwidth]{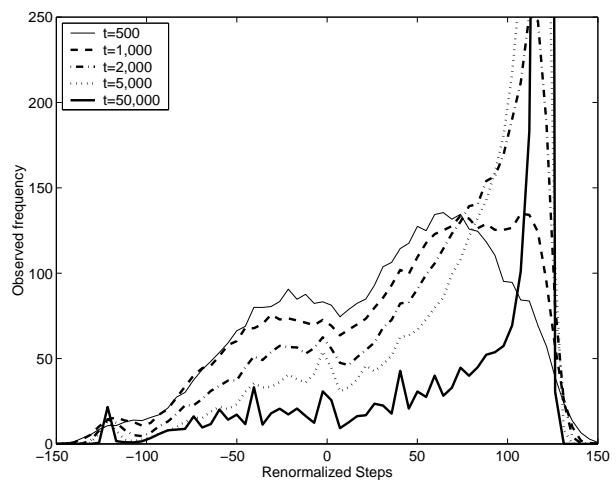}
 \caption{Average distribution of opinions of the agents after different number of updates, with $t$ measuring thousands of individual updates. It corresponds to the random network with 8 neighbors on average.}
 \end{figure} 
 
 Figure~4 shows the same analysis for the case where we have converted a 8 neighbors case to a random graph, with a rewiring probability of $\lambda=1.0$, that is, a random network with 8 neighbors on average. As we had seen in Figure~2, this case showed the appearance of a strong consensus, but a small minority of dissenters remained. In order to compare the shape of the curves, a renormalization of the opinions is necessary. Basically, for a lattice where the actions are stable, each agent still updates its internal opinion in the direction of the majority of its neighbors, following a random walk. That means that follows a random walk and the opinion moves, in average, with a constant velocity away from the moderate opinions, as discussed in the first article~\cite{martins08a}. Therefore, if one doubles the number of interactions, the opinions will tend to be at a distance away from flipping that is doubled and, for easy of comparison, they need to be renormalized. The results in Figure~4 are treated that way. 
  
\begin{figure}[bbb]\label{fig:regular}
 \includegraphics[width=0.45\textwidth]{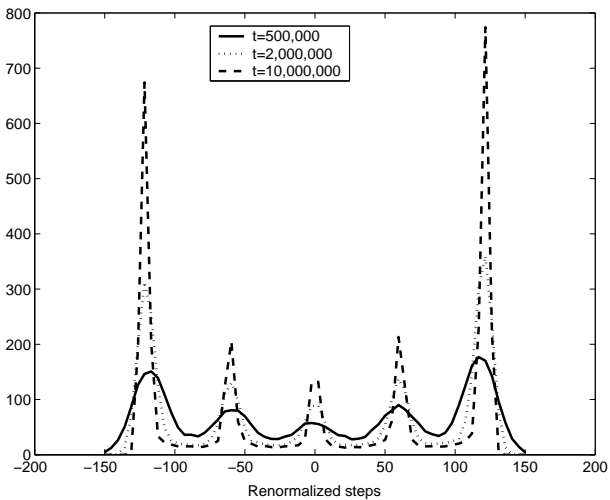}
 \caption{Average distribution of opinions of the agents after different number of updates, with $t$ measuring thousands of individual updates. It corresponds to a regular network with 4 neighbours (von Neumann neighborhood).}
 \end{figure} 
 
 Figure~5 shows the evolution for a regular lattice with four neighbors and no rewiring, for comparison. Here, we can see that, not only the choices become stable, as it had been observed before~\cite{martins08a}, but the shapes also tend to a stable configuration, with each peak becoming more and more important. In the long run, the system seems to tend to a sum of five delta functions and that is due to the fact, discussed in the previous paper, that each agent opinion basically follows a random walk after the system has reached a stable state.
 
Notice that, as the number of interactions between the agents increases, the majority peak becomes more and more important and the rest of the opinions tend to diminish. However, this is not true for the minority peak. It remains small, but it also gets better defined. That behavior is consistently observed even after 50,000,000 of individual interactions, corresponding to a little more than 12,000 opinion updates in average, for each agent. Notice that the real, non-renormalized, opinions are actually becoming more distant as time passes. Actually, as in every case observed, the peaks move away from the center with a constant speed. The same effect has been observed for regular lattices, where the choices became stable and each observed peak only became more well defined after more time\cite{martins08a}.

It is interesting to compare these results with those of bounded confidence models~\cite{deffuantetal00,hegselmannkrause02,amblarddeffuant04,deffuant06}. In bounded confidence models, the agents could influence each other if the distance between their opinions was smaller than a certain threshold value. This caused the final population state to either converge to a consensus where everybody agreed, or to a number of different final opinions, that got larger as the threshold became smaller. In Figure~5, we see an apparently similar result with 5 possibilities surviving and, as we will see in the next Section, under some circumstances, consensus may emerge. However, one should remember that the peaks in Figure~5 do not correspond to fixed values of $v$, since $v$ is always increasing. They are shown in the same position due to the renormalization. And, from the point of view of what is observed, each agent still makes a binary choice and the agents never express the values they associate with each choice to anyone else. The most extreme peak agrees with the less extreme one, in their choices, as long as they have the same sign for $v$.

The effects of spatial structure in models of bounded confidence have been studied. Weisbuch et al~\cite{weisbuchetal05} describe how a few initial extremists, with very small threshold for their opinions, can cause the spread of their extreme opinions. Spatial structure seemed to have an effect that allowed centrists to survive better and different descriptions were observed depending on the value of the threshold for centrists. Deffuant~\cite{deffuant06} observed that clusters of moderates appear more easily in networks where the individuals tend to be closer to each other. Here, we have seen that this seems to be the case, since the extremist peaks became less important when the network tends to a random graph. This effect was particularly more noticeable when the agents had, in average, more links, confirming Deffuant's observation. However, important differences were observed in the CODA model, since no cluster of moderates is observed.

\section{Introducing Mobility}

In every scenario investigated in the previous Section, the neighborhood of each agent was determined in the beginning of the simulation and it was never changed after that. However, sometimes, real people do change the place they live in or their influences. In order to study that, some mobility must be introduced in the problem.

\begin{figure}[bbb]
 \includegraphics[width=0.5\textwidth]{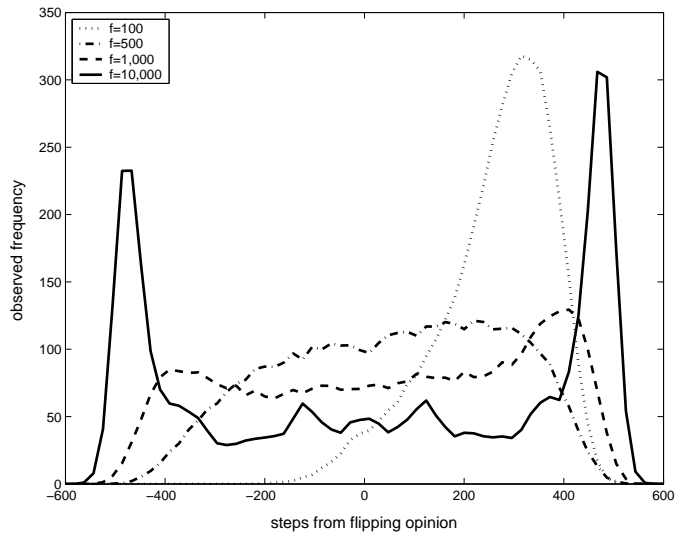}
 \label{fig:mobility}
 \caption{Average distribution of the opinions of agents in a 64x64 square lattice after 2,000,000 of opinion updates (about 488 average updates per agent). The simulations show the results for a regular square lattice with a von Neumann neighborhood (4 neighbors) when the agents change their position in the network at random, after $f$ opinion updates.}
\end{figure}

Figure~6 shows such scenario. There, the effects of agents changing their spatial position in the lattice were investigated. The change was performed by randomly choosing two agents to swap places, once every $f$ opinion updates. Therefore, $f$ is related to how many times an agent update its opinion, in average, before it changes its position. Since two agents change their position after $f$ individual opinion updates, $f/2$ is the correct factor to be used for comparisons (average number of times an agent updates its opinion before it changes place). Migration caused by opinion differences was studied before, under the context of active Brownian particles~\cite{schweitzerholyst00}.  In that paper, migration was caused by the differences in opinions. Here, migration is assumed to be an external phenomenon, independent of the opinions and, therefore, considered completely random. Observe that, for a large value of $f$ ($f/2=5,000$ opinion updates per moving), mobility seems to have almost no effect. But it becomes increasingly more important as $f$ decreases. Between $f/2=250$ and $500$ opinion updates per moving, centrists become far more common, even though extremists are still observed. Compared to network effects, however, the decrease in extremism is significant.

\begin{figure}[bbb]\label{fig:averageop}
 \includegraphics[width=0.45\textwidth]{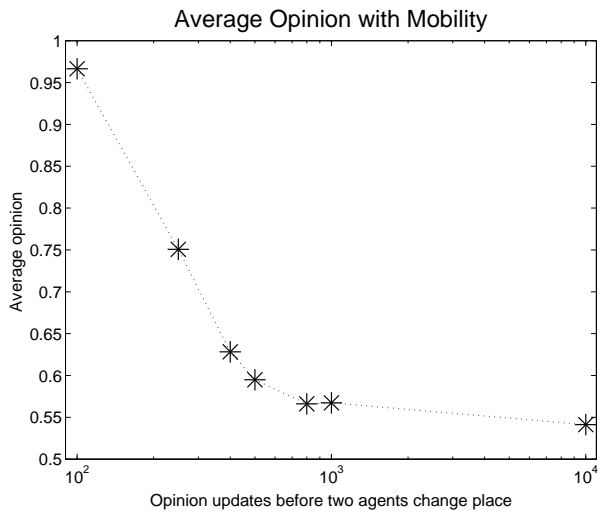}
 \caption{Average percentual opinion in favor of the majority opinion, after 2,000,000 opinion updates, as a function of the number of opinion updates $f$ performed between the exchange of places of two agents.}
 \end{figure}

However, as $f$ gets even smaller, a different effect can be observed. Instead of the appearance of a majority of agents with moderate opinions, we observe the appearance of a clear majority, where most people have the same opinion (in the long run, the system tends to complete consensus). This happens due to the fact that, given the high mobility, after a while, everyone is influenced by everyone. For large values of $f$, this long run might take so long to happen that this effect will not be observed in a reasonable amount of time. The average effect of mobility can be seen clearly in Figure~7, where the average proportion of people supporting the majority opinion is shown as a function of $f$. Notice that, as $f$ becomes smaller, a change in regime happens and, instead of a split population, one observes the appearance of a clear majority.

Another interesting aspect of the model is its temporal evolution. In the original article~\cite{martins08a}, we have seen that the configuration of the observed opinions for regular lattices becomes stable much earlier than after about 500 updates per agent, as used in all simulations discussed so far, as we can see in Figure~5.

\begin{figure}[bbb]\label{fig:distmobf100}
 \includegraphics[width=0.45\textwidth]{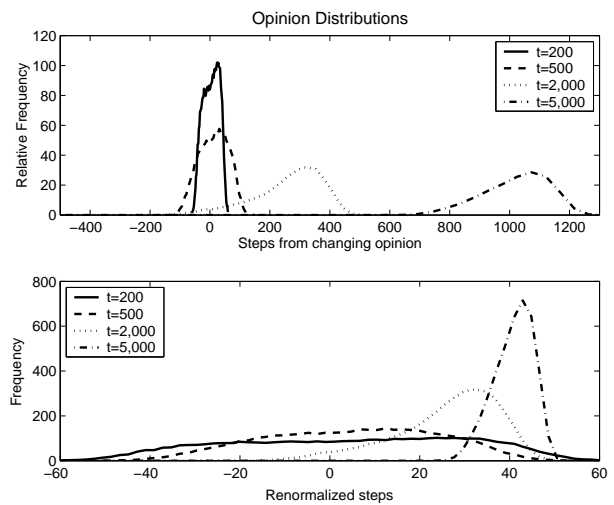}
 \caption{Average percentual opinion for the opinion of the agents after different number of updates, with $t$ measuring thousands of individual updates. Both graphs correspond to the originally regular neighbors lattice (von Neumann neighborhood), with mobility happening at a rate given by $f=100$. The upper graph shows the actual number of steps away from flipping the opinion, while the lower one shows renormalized steps, so that the peaks will be observed in the same region.}
 \end{figure}
 
However, since we have observed here a few situations where consensus emerges, it makes sense to analyze the temporal evolution of the distributions for those cases. Figure~7 shows the distribution of continuous opinions for four different number $t$ of individual updates, in the mobility case, for $f=100$. The upper graph shows the actual opinions and we can see that, initially, they are distributed around 0, with a very weak tendency for the appearance of two peaks. As one of the two opinions become more important, the tendency towards consensus become stronger than the tendency to polarization and the population starts moving as a whole to the positive values (majority). The lower graph shows exactly the same case, but with the steps away from $v=0$ renormalized by the number of interactions so that extremist positions can happen at basically the same position. It shows more clearly the tendency for the consensus, as the peak becomes higher. 

It is interesting to notice that $f$ seems to be defining a time scale. For $N^2$ agents in the simulations, since two of them are changed each time agents change position, $N^2$ agents get changed after $N^2 f/2$ opinion updates. Assuming that half of the network, initially, supports each position, it should take an average time of $N^2 f$ updates before an extremist agent is put in contact with a majority of agents with the opposite opinion. In Figure~7, $N^2 f \approx 400,000$. We can see that, after 500,000 a majority is starting to take shape, while after 2,000,000 updates, the system has almost reached full consensus. In order to investigate this hypothesis, the case $f=10,000$ was simulated using this scale. After 50,000,000 updates, the extremists have indeed disappeared and the same shape as that observed for $f=500$ after 2,000,000 of observations emerge. That is a central distribution around zero, with two small peaks, originated from the extremist peaks tending to less extreme positions. After that, the majority turns into a consensus and the behavior observed in Figure~8 emerges. 

It is easy to understand how that happens, since, in the long run, regardless of $f$, all agents will have an opportunity to interact with every other agent and the model becomes basically a mean field model. In that case, an opinion with a small advantage will always tend to dominate, since each agent will have a smaller than half chance to move towards that direction. This will turn the small majority into a larger one and the convergence will then becomes faster. That means that the value of $f$ is only important for the dynamics in smaller time scales.

\section{Conclusion}

These simulated results suggest that increasing contact between different opinions tend to make them less extreme. Social extremism, defined as the co-existence of two positions defended by extremists, seems to be, at least partially, the consequence of little interaction between people with different ideas. The amount of extremism observed seems to be somehow related to the structure of the society, as different networks (small worlds, diagonal neighbors, etc.) produced a different quantity of extremists. But it seems that, for any rigid society structure, some extremism is always observed. Finally, we see that the extremism problem can become far less important in societies where the mobility of its agents is above a certain threshold. Given enough time, the mobility will always cause the consensus to emerge, but this might take too long and too many interactions to be useful. Therefore, efforts to reduce such mobility can have important negative impacts in the diminishing of extremism.

\begin{acknowledgments}
The author would like to thank Renato Vicente for discussing some of the ideas here presented and for reviewing the first version of this paper. The author would also like to thank Funda\c{c}\~ao de Amparo \`a Pesquisa do Estado de S\~ao Paulo (FAPESP), for the support to this work.
\end{acknowledgments}

\bibliography{extremism}

\end{document}